\documentstyle[12pt]{article}

\input epsf

\newcommand{\mpl}{m_{\rm Pl}}
\newcommand{\nt}{\widetilde{n}}
\newcommand{\Ol}{\Omega_\Lambda}
\def\degrees{\hbox{${}^\circ$\hskip-3pt .}}
\def\spose#1{\hbox to 0pt{#1\hss}}
\def\simlt{\mathrel{\spose{\lower 3pt\hbox{$\mathchar"218$}}
	\raise 2.0pt\hbox{$\mathchar"13C$}}}

\begin{document}
\pagestyle{empty}
\begin{center}
\bigskip

\rightline{FERMILAB--Pub--95/405-A}
\rightline{astro-ph/9512155}
\rightline{submitted to {\it Physical Review D}}

\vspace{.2in}
{\Large \bf Dependence of Inflationary Reconstruction upon \\
\bigskip
Cosmological Parameters}
\bigskip

\vspace{.2in}
Michael S. Turner$^{1,2,3}$ and Martin White$^{1,2}$\\

\vspace{.2in}
{\it $^1$NASA/Fermilab Astrophysics Center\\
Fermi National Accelerator Laboratory, Batavia, IL~~60510-0500}\\

\vspace{0.1in}
{\it $^2$Department of Astronomy \& Astrophysics\\
Enrico Fermi Institute, The University of Chicago, Chicago, IL~~60637-1433}\\

\vspace{0.1in}
{\it $^3$Department of Physics\\
Enrico Fermi Institute, The University of Chicago, Chicago, IL~~60637-1433}\\

\end{center}

\vspace{.3in}
\centerline{\bf ABSTRACT}
\bigskip
The inflationary potential and its derivatives determine the spectrum of
scalar and tensor metric perturbations that arise from quantum fluctuations
during inflation.  The CBR anisotropy offers a promising means of determining
the spectra of metric perturbations and thereby a means of constraining the
inflationary potential.  The relation between the metric perturbations and
CBR anisotropy depends upon cosmological parameters -- most notably the
possibility of a cosmological constant.  Motivated by some observational
evidence for a cosmological constant (large-scale structure, cluster-baryon
fraction, measurements of the Hubble constant and age of the Universe) we
derive the reconstruction equations and consistency relation to second order
in the presence of a cosmological constant.
We also clarify previous notation and discuss alternative schemes for
reconstruction.

\newpage
\pagestyle{plain}
\setcounter{page}{1}
\newpage

\section{Introduction}

Inflation gives rise to nearly scale-invariant scalar (or density) and tensor
(or gravity-wave) metric perturbations which are excited by quantum
fluctuations in the inflaton field and in the metric itself \cite{perts} and are
determined by the inflationary potential and its derivatives \cite{lidlyt}.
Measurements of the scalar and tensor metric perturbations permit partial
reconstruction of the inflationary potential \cite{reconstruct}.
Both the scalar and tensor perturbations give rise to temperature anisotropies
in the cosmic background radiation (CBR) \cite{cbranis}, and precise
measurements of CBR anisotropy on angular scales from $0\degrees1$ to
$100^\circ$ (multipole numbers $\ell=2-1000$) probably offer the most
promising means of determining the metric perturbations.

Copeland and his collaborators have emphasized the underlying relationship
that exists between the inflationary potential and the power spectra
describing the metric perturbations \cite{ckll} (`$k$'-space reconstruction).
These relations are independent of present cosmological parameters
(e.g., Hubble constant, baryon density, cosmological constant, and ionization
history of the Universe).  On the other hand, Turner has emphasized that realizing reconstruction
in practice requires connecting the potential and its first few derivatives to
a handful of observables, e.g., the scalar and tensor contributions to the
quadrupole CBR anisotropy ($S$ and $T$) and the power-law spectral indices
that characterize the scalar and tensor power spectra
($n$ and $n_T$) (`$\ell$'-space reconstruction).
However, relating the power spectrum to CBR anisotropy necessarily brings in
these cosmological parameters \cite{turner}.

As we shall discuss, the only significant dependence that arises in going from
$k$-space to $\ell$-space is from a possible cosmological constant, and the
purpose of this paper is to quantify that dependence.
At the moment there is some motivation for considering a cosmological constant:
the cold dark matter model with a cosmological constant provides a
good fit to all the observational data
(large-scale structure, measurements of the Hubble constant and cluster baryon
fraction, and age of the Universe) \cite{lcdm}.
In addition to calculating the dependence of the reconstruction and consistency
equations upon the value of the cosmological constant and showing that the
dependence upon other cosmological quantities is insignificant, we will clarify
previous notation and normalization conventions and discuss alternative
reconstruction strategies.

\section{Perturbative Reconstruction}

The program of perturbative reconstruction is spelled out in
Refs.~\cite{perturbative}.  The basic idea is to express a handful of
observables -- e.g., $S$, $T$, $n$, $n_T$, and $dn/d\ln k$ -- in terms of the
derivatives of the inflationary potential, evaluated at some convenient point
(here denoted by `*').  The perturbation expansion is in the deviation from
exactly exponential inflation and exact scale invariance, quantified by the
derivatives of the potential,
or more conveniently in terms of the spectral indices $n_T$ and $\nt\equiv n-1$.
In the scale-invariant limit $n_T=\nt=0$ and all the derivatives of the
potential vanish.  

\subsection{Notation and lowest-order reconstruction}

Let's begin at lowest order.  To lowest order (in $n_T$ and $\nt$) the power
spectra of scalar and tensor perturbations are described by
\begin{eqnarray} \label{indices}
\nt & = & -{3\over 8\pi}\left({\mpl V^\prime_* \over V_*} \right)^2
        + {1\over 4\pi}\left( {\mpl^2V_*^{\prime\prime}\over V_*} \right) \\
n_T & = & -{1\over 8\pi} \left(
        {\mpl V^\prime_* \over V_*} \right)^2
\end{eqnarray}
\begin{eqnarray}
P(k) & = & {1024\pi^3\over75}\,{k \over H_0^4}\,
        \left({g(\Omega_0)\over\Omega_0}\right)^2
        {V_*^3\over\mpl^6{V_*^\prime}^2}
        \left[k/k_*\right]^{\nt} T^2(k) \label{ps} \\
P_T(k) & = & {8\over 3\pi}{V_*\over \mpl^4}\left[k/k_*\right]^{n_T} T_T^2(k)
  \label{pg}
\end{eqnarray}
where $V(\phi )$ is the inflationary potential, prime denotes $d/d\phi$,
and $V_*=V(\phi_*)$, etc.

The factor $[g(\Omega_0)/\Omega_0]^2$ in Eq.~(\ref{ps}) takes into
account the growth of the perturbations and the relation between the density
and curvature perturbations, with $g(\Omega_0)$ well fit by \cite{CPT}
\begin{equation}
g(\Omega) = {5\over 2} \Omega \left[ {1\over 70} + {209\Omega\over 140}
        - {\Omega^2\over 140} + \Omega^{4/7} \right]^{-1}
\end{equation}
where $\Omega_0$ is the matter density (cold dark matter + baryons).
The functions $T(k)$ and $T_T(k)$ are the ``transfer functions'' which describe
the cosmological evolution of the modes that arises due to the transition of
the Universe from an early radiation-dominated epoch to a matter-dominated
epoch, and are defined so that $T(k), T_T(k)\rightarrow1$ for $k\rightarrow 0$.
The transfer functions together with the growth factor for scalar perturbations
take ``primordial'' spectra to presently ``observed'' spectra.
The scalar transfer function can be fit by \cite{Peebles,EfsBonWhi}
\begin{equation}
T(k) = \left[ 1+\bigl( ak+(bk)^{3/2}+(ck)^2\bigr)^\nu\right]^{-1/\nu}
\end{equation}
with $a=(6.4/\Gamma)$Mpc, $b=(3.0/\Gamma)$Mpc, $c=(1.7/\Gamma)$Mpc
and $\nu=1.13$.  Here $\Gamma\simeq\Omega_0h$ is a measure of the size of
the horizon at matter-radiation equality.
In the absence of a cosmological constant, the gravity-wave transfer function
can be written as \cite{TurWhiLid}
\begin{equation}
T_T(k) = {3 j_1(k\tau_0) \over k\tau_0 }{\cal T}(k)
\end{equation}
where $j_1(x)$ is the spherical bessel function of the first order and
${\cal T}(k)$ is a factor analogous to $T(k)$ and is given in
Ref.~\cite{TurWhiLid}. For the modes of most interest, those that
enter the horizon during matter domination, ${\cal T}(k)\approx1$.

It is convenient to rewrite the scalar and tensor spectra in terms of
quantities $A_S^2(k)$ and $A^2_T(k)$ whose only $k-$dependence arises from a
deviation from scale invariance\footnote{Copeland et al.~have introduced
several definitions of $A_S^2$ and $A_T^2$ (also denoted as $A_G^2$); we shall
use the definitions given in Ref.~\cite{rmp}.  We note that the relationship
between $A_i^2$ and the power spectra given in the Appendix of
Ref.~\cite{arlmst} has errant factors of $k_{50}^{n-1}$ and $k_{50}^{n_T}$ as
well as not using the current definitions of the $A_i^2$.  Copeland et al.~have
not explicitly discussed the definitions of $A_i^2$ for $\Omega_\Lambda\ne0$;
we define without the $\Omega_0$ dependence that arises from the growth of
perturbations and relating density and curvature perturbations.}
and which have a simple physical interpretation,
\begin{eqnarray}
P(k) & \equiv & {2\pi^2 \over H_0^4}\,k\ A_S^2(k)
       \left({g(\Omega_0)\over\Omega_0}\right)^2\,T^2(k)\\
P_T(k) & \equiv & {25\over 4\pi}A_T^2(k)\,T_T^2(k)
\end{eqnarray}
where to lowest order in $\nt$,  $n_T$
\begin{eqnarray}
A_S^2(k) & = & {512\pi\over 75} [k/k_*]^{\nt} \,{V_*^3\over
        \mpl^6{V_*^\prime}^2} \\
A_T^2(k) & = & {32\over 75}[k/k_*]^{n_T}\,{V_*\over \mpl^4}
\end{eqnarray}

For the scalar perturbations, $A_S^2(k=H_0)$ is the present contribution of
this mode to the {\it rms} mass fluctuation per logarithmic interval in $k$
(in the absence of a cosmological constant):
\begin{equation}
A_S^2(k=H_0) = \Delta^2(H_0) \equiv
\left. {k^3\over 2\pi^2}P(k)\right|_{k=H_0}
\end{equation}

The quantity $A_T^2(k=H_0)$ is related to the present energy density in
long-wavelength, inflation-produced gravitational waves (in the
absence of a cosmological constant):
\begin{equation}
{d\log\Omega_{\rm GW} \over d\ln k} =
        {75\over32}A_T^2(k)\left({k\over H_0}\right)^{-2},
\end{equation}
valid for $k\ll k_{\rm EQ}\sim200H_0$.

In Eqs.~(\ref{ps},\ref{pg}) the expansion point $\phi_*$ is the point about
which the power-law indices and all the derivatives of the potential are
evaluated.  It is defined by the fact that the comoving scale $k_*$ crossed
outside the horizon during inflation when $\phi=\phi_*$; given the details of
inflation (reheat temperature and so on) it is straightforward to relate
$\phi_*$ to the number of e-foldings before the end of inflation and/or to
$k_*$.  For lowest-order reconstruction $\phi_*$ is irrelevant as any
dependence upon it involves higher-order corrections in $n_T$ and $\nt$.
For second-order reconstruction, the choice of $k_*$ is important.  It will
prove very convenient to choose $k_*=H_0$; later we will discuss the dependence
of the reconstruction equations upon $k_*$.

In order to practically implement reconstruction the power spectra
must be related to observables.  At lowest order the natural
set of observables is $S$, $T$, $\nt$, and $n_T$.  Since $\nt$ and
$n_T$ are already given in terms of the potential, it only remains
to relate $S$ and $T$ to $A_S^2(k_*)$ and $A_T^2(k_*)$.
The variance of the multipole moments of the expansion of the
CBR temperature field are integrals of the power spectra times
kernels which depend on the cosmological parameters
($h$, $\Omega_Bh^2$, $\Ol$) and the ionization history
of the Universe. These integrals introduce the dependence of the
reconstruction equations upon cosmological parameters.
The dependence upon all of these except $\Ol$ is very weak
(less than 1\% for sensible variations in $\Omega_Bh^2$ and 4\% for
sensible variations in $h$).

The ``Rosetta Stone'' relations for lowest-order reconstruction,
which take $k$-space equations to $\ell$-space equations, are:
\begin{eqnarray}
S\equiv{5C_2^S\over4\pi} &=& 0.10f_S^{(0)}(\Ol) A_S^2(k_*) \\
T\equiv{5C_2^T\over4\pi} &=& 1.4\phantom{2} f_T^{(0)}(\Ol) A_T^2(k_*)
\end{eqnarray}
where we have followed conventional practice and expanded the two-point
function of the CBR temperature perturbations in Legendre polynomials
\begin{equation}
\left\langle {\Delta T\over T}(\hat{x}_1)\,
{\Delta T\over T}(\hat{x}_2)\right\rangle\equiv
{1\over 4\pi}\sum_\ell\,(2\ell+1)\,C_\ell\,P_\ell(\hat{x}_1\cdot\hat{x}_2)
\end{equation}
where brackets denote the average over the sky.
The functions $f_S^{(0)}(\Ol)$ and $f_T^{(0)}(\Ol)$ quantify the dependence of
reconstruction upon the cosmological parameter $\Ol=1-\Omega_0$.
We have evaluated them numerically (with $h=0.75$ and $\Omega_Bh^2=0.0125$)
and normalized them such that all expressions have their familiar values with
$f_i^{(0)}\simeq1$.  The functions and their ratio are shown in Fig.~1;
they are well fit by quadratics over the range $0.0\le\Ol<0.8$:
\begin{eqnarray}
f_S^{(0)} & = & 1.04 - 0.82\Ol + 2\Ol^2\\
f_T^{(0)} & = & 1.0\phantom{4} - 0.03\Ol - 0.1\Ol^2
\end{eqnarray}
The correction to the familiar scalar relation in the $\Ol=0$ limit
(i.e.,~$f_S^{(0)}(0)\ne1$) arises from including the integrated Sachs-Wolfe
effect, due to the decay of the potentials near matter-radiation equality
(see also Fig.~2).

Using these relations in place of the scalar and tensor power spectra, the
lowest-order reconstruction equations and consistency relation follow directly:
\begin{eqnarray}
{V_*\over\mpl^4} &=& 1.65\, T/f_T^{(0)} \\
{V_*^\prime\over\mpl^3} &=& \pm 8.3\,\sqrt{-n_T}\,T/{f_T^{(0)}} \\
{V_*^{\prime\prime}\over\mpl^2} &=& 21\,(\nt - 3n_T)\,T/{f_T^{(0)}} \\
n_T & = & -{1\over 7}{f_S^{(0)} \over f_T^{(0)}}{T \over S}
\end{eqnarray}
where the sign of $V_*^\prime$ is indeterminate as it
can be changed by taking $\phi$ to $-\phi$.
The final expression is the consistency relation that arises since the four
observables can be expressed in terms of three properties of the potential.
The familiar factor of 1/7 is modified by ratio of $f_S^{(0)}/f_T^{(0)}$,
introduced in Ref.~\cite{knox}.  In practice $n_T$ is likely to
be difficult to measure, and so the consistency relation can
be used to eliminate $n_T$ in the expressions for $V_*^\prime$ and
$V_*^{\prime\prime}$.

\subsection{Second-order reconstruction}

Including the $\Ol$ dependence in second-order reconstruction is
in principle as easy as it was in lowest-order reconstruction.
(Second-order refers to including the order $\nt$ and $n_T$ corrections
to the reconstruction and consistency equations.)  However the strategy
is slightly different because while there are second-order
expressions for the power spectra, cf.~Ref.~\cite{arlmst},\footnote{We
will not need these expressions; in any case, the second-order
corrections are multiplicative factors of $1+7n_T/6 + (-7/3 +\ln 2 +
\gamma )\nt$ to $A_S^2$ and of $1+(-7/6 +\ln 2 +\gamma )n_T$ to $A_T^2$.}
similar explicit expressions for the spectral indices do not exist.
In addition, another observable is needed; the plausible candidate is
the ``running'' of the scalar spectral index \cite{komst},
$dn/d\ln k$, which is ${\cal O}(\nt^2 ,n_T^2)$.

Reconstruction proceeds from $k$-space expressions relating
the inflationary potential and its derivatives at $\phi_*$ to
$A_T^2(k_*)$ and $A_S^2(k_*)$, and follows the $\Ol=0$
case done in Ref.~\cite{arlmst}.  The key $k$-space equations are
\begin{eqnarray}
{V_*\over \mpl^4} & = &
        {75 \over 32}A_T^2\left[ 1+0.21{A_T^2\over A_S^2} \right] \\
{V_*^\prime \over \mpl^3} & = & -{75\sqrt{\pi}\over 8}
        {A_T^3\over A_S} \left[
        1- 0.85{A_T^2 \over A_S^2} -0.53\nt\right] \\
{V_*^{\prime\prime}\over \mpl^2} & = & {25\pi\over 4}
        A_T^2 \left[ \nt + 6{A_T^2\over
        A_S^2} -16{A_T^4\over A_S^4} -{\nt^2\over 6}
        -9.8\nt{A_T^2\over A_S^2}+1.1{d\nt\over d\ln k}  \right]\\
{V_*^{\prime\prime\prime}\over\mpl} & = & \pm 4\pi\sqrt{-8\pi n_T}
        \left[ {d\nt/d\ln k\over n_T} -6n_T +4\nt \right] {V_*\over\mpl^4} \\
{A_T^2\over A_S^2} & = & -{n_T\over 2}\left[ 1 - {n_T\over 2} + \nt\right]
\end{eqnarray}
where for simplicity the arguments of $A_S(k_*)$ and $A_T(k_*)$ have been
omitted.  Note too that the spectral indices and the
derivative of the scalar spectral index are also evaluated
at $k_*$.  These equations are Eqns.~(3.4), (3.6), and (3.15) of
Ref.~\cite{ckll2} and Eqns.~(39) and (46) of
Ref.~\cite{arlmst}, as modified to be consistent with the definitions
of $A_S^2$ and $A_T^2$ in Ref.~\cite{rmp}.  The last equation is the
second-order consistency equation.  It can be used to eliminate the
factors of $A_T^2/A_S^2$ in the first three equations.

Once again, the key to going from $k$-space equations to $\ell$-space
equations is relating $A_T^2(k_*)$ and $A_S^2(k_*)$ to the CBR observables
$T$ and $S$.  At second order, the order $\nt$ and $n_T$ corrections must
be taken into account.  The second-order ``Rosetta Stone'' equations
are given by
\begin{eqnarray}
S & = & 0.10 f_S^{(0)}(\Ol)[1+f_S^{(1)}(\Ol)\nt_{\phantom{T}}] A_S^2(k_*) \\
T & = & 1.4\phantom{2} f_T^{(0)}(\Ol) [1 + f_T^{(1)}(\Ol)n_T  ]A_T^2(k_*)
\end{eqnarray}
where $f_i^{(0)}(\Ol)$ are the same functions are in the previous Section
and second-order expressions for $A_i(k_*)$ must be used.
The functions $f_i^{(1)}(\Ol )$ quantify the $\Ol$ dependence of the
second-order corrections that arise from relating the $A_i^2$ to
$S$ and $T$.  They depend upon the ``pivot point'' $k_*$ and for
$k_*=H_0$, they can be accurately fit by:
\begin{eqnarray}
f_S^{(1)} &=& 0.45-0.51\Ol+1.04\Ol^2-0.14\Ol^3  \\
f_T^{(1)} &=& 0.58-0.50\Ol+0.31\Ol^2-0.88\Ol^3
\end{eqnarray}

Concerning the pivot-point dependence of $f_i^{(1)}(\Ol)$; using the fact that
$A_S^2(k)\propto [k/k_*]^{\nt}$ and $A_T^2(k)\propto [k/k_*]^{n_T}$ it is
simple to show that under the change $k_*\rightarrow k_*^\prime$,
$f_i^{(1)}\rightarrow f_i^{(1)}+\ln(k_*/k_*^\prime )$.  We note that
changing the pivot point does not affect the form of higher-order corrections,
i.e., the values of $f_i^{(j)}$ for $j=2,\cdots$.

The $\ell$-space reconstruction and consistency equations now follow
from the $k$-space equation through use of the ``Rosetta Stone'' equations:
\begin{eqnarray}
{V_*\over \mpl^4} &=&
        1.65 \left[ 1. - (f_T^{(1)}+0.1)n_T \right]\,T/f_T^{(0)} \\
{V_*^\prime\over \mpl^3} &=&
        \pm 8.3\sqrt{-n_T} \left[1. -(f_T^{(1)}-0.18)n_T-0.03\nt \right]\,
        T/f_T^{(0)} \\
{V_*^{\prime\prime}\over \mpl^2} &=&
	21\left[(\nt-3n_T)+(3f_T^{(1)}-2.6)n_T^2+(1.9-f_T^{(1)})n_T\nt
	\vphantom{\int}\right.\\
&& \left. -0.2\nt^2 +1.1{d\nt\over d\ln k}\right]\,T/f_T^{(0)}\\
{V_*^{\prime\prime\prime}\over\mpl} &=&
	\pm 104\sqrt{-n_T}
        \left[ {d\nt/d\ln k \over n_T} -6n_T + 4\nt \right]\,T/f_T^{(0)} \\
n_T & = & -{1\over 7}{f_S^{(0)}\over f_T^{(0)}}{T\over S}
        \left[ 1+{\textstyle{1\over 7}}(f_T^{(1)}-{\textstyle{1\over 2}})
        (f_S^{(0)}/f_T^{(0)}){T\over S}
        +(f_S^{(1)}-1)\nt \right]
\end{eqnarray}
While the signs of $V_*^\prime$ and $V_*^{\prime\prime\prime}$ are
arbitrary, the relative sign is not.  By using the consistency equation the
factors of $n_T$ (which is likely to be
very difficult to measure) can be eliminated in favor of $T/S$.

Finally, we note that the previous results for
$\Ol = 0$ in Ref.~\cite{arlmst} can be recovered by substituting
$f_T^{(0)}=f_S^{(0)}=1,$ $f_T^{(1)}=1.3$, and $f_S^{(1)}=1.15$.
(In Ref.~\cite{arlmst} the
pivot point $k_* = H_0/2$, so that $\ln 2$ must be added
to the $f_i^{(1)}$ defined above.)

\subsection{Alternative schemes}

The goal of perturbative reconstruction is to use data, most likely
measurements of CBR anisotropy, to infer the value of the inflationary
potential and its first few derivatives at $\phi_*$.  To achieve this
goal, one needs to pick a set of observables and then relate the power
spectra to these observables.  At second order, the pivot point $k_*$ also
comes into play.

The spectral indices $\nt$, $n_T$, and $dn/d\ln k$ (for second order) are
obvious choices for the observables (though in practice one will probably
wish to use the consistency equation to eliminate $n_T$).  The quantities
$S$ and $T$ are sensible choices as they:
(i) serve to normalize the scalar and tensor contributions to CBR anisotropy;
(ii) are easy to extract from CBR measurements; and
(iii) are relatively insensitive to all the cosmological parameters except $\Ol$.
If one uses $S$ and $T$ then it is also sensible to select the pivot point
$k_*=H_0$, which minimizes the dependence of $S$ and $T$ upon $n_T$ and
$\nt$ since the dominant contribution to $S$ and $T$ comes from modes with
$k\sim H_0$.

On the other hand, since the multipoles that will have the most leverage in
determining $\nt$ (and $dn/d\ln k$) are $\ell\sim30-300$, it might be more
useful to choose $k_*\sim (30-100)H_0/2$ (recall, $\nt$ and $dn/d\ln k$ are
evaluated at $k=k_*$).  However, the higher multipoles are more sensitive to
the cosmological parameters (e.g., $h$ and $\Omega_Bh^2$).

In any case, it is a simple matter to substitute other multipoles
for $S$ and $T$.  For example, consider
\begin{eqnarray}
S_{30} & = & {61 C_{30}^S \over 4\pi}\\
T_{30} & = & {61 C_{30}^T \over 4\pi}
\end{eqnarray}
Writing the ``Rosetta Stone'' equations in precisely
the same form as before,
\begin{eqnarray}
S_{30} &=& 0.10 f_S^{(0)}(\Ol)[1+f_S^{(1)}(\Ol)\nt_{\phantom{T}}] A_S^2(k_*) \\
T_{30} &=& 1.4\phantom{2} f_T^{(0)}(\Ol)[1+f_T^{(1)}(\Ol)n_T  ]A_T^2(k_*)
\end{eqnarray}
the form of the $\ell$-space reconstruction equations and consistency relation
are unchanged (except $T\rightarrow T_{30}$ and $S\rightarrow
S_{30}$).  It must of course be remembered that $f_i^{(0)}(\Ol )$ and
$f_i^{(1)}(\Ol )$ are completely different functions which also have
significant dependence upon other cosmological parameters.  Taking
$k_*=20H_0$, $\Omega_Bh^2=0.0125$ and $h=0.75$, the $\Ol$ dependence can be
fit by,
\begin{eqnarray}
f_S^{(0)}(\Ol) & = & \phantom{-}0.11 - 0.02\Ol + 0.07\Ol^2  \\
f_T^{(0)}(\Ol) & = & \phantom{-}0.08 - 0.00\Ol + 0.01\Ol^2   \\
f_S^{(1)}(\Ol) & = &
\phantom{-}0.25 - 0.35\Ol + 0.07\Ol^2 - 0.52\Ol^3 - \ln(k_*/20H_0) \\
f_T^{(1)}(\Ol) & = &-0.11 - 0.54\Ol + 0.37\Ol^2 - 0.93\Ol^3 - \ln(k_*/20H_0)
\end{eqnarray}

\subsection{Power-law inflation: an exact result}

If the inflaton potential is an exponential,
\begin{equation}
V(\phi ) = V_0 \exp [-\sqrt{16\pi/p}\ \phi /\mpl ],
\end{equation}
the growth of the scale factor
during inflation is precisely a power law, $R(t) \propto t^p$, and it is
possible to solve for the perturbation spectra exactly \cite{PLIspectra}.
In this case the only parameter to be determined is the Hubble constant
during inflation ($=H_*$) when the mode $k_*$ crossed outside the horizon.

For power-law inflation the solution of the equation of motion
(the massless Klein-Gordon equation) for fluctuations in the inflaton and the
gravitational fields is a Hankel function.
For modes that are well outside the horizon at the end of inflation
(all those of astrophysical interest are), matching values of the field
and its first derivative at the end of inflation allows one to
calculate the Bogoliubov coefficients relating the creation and annihilation
operators describing the quantum field before and after the end of inflation
(see e.g., Ref.~\cite{White} and references therein).  From
these one can calculate the two-point function of the
(classical, random) field at the present, and $P(k)$ and $P_T(k)$.
They are exact power laws with spectral indices $\nt=n_T=-2/(p-1)$,
and\footnote{Exponential inflation is also analyzed in Ref.~\cite{arlmst};
the second-order correction to the power spectra, given in Eq.~(52),
is missing a factor of $1/\sqrt{1-1/3p}$.  This improves significantly
the accuracy of the reconstruction of an exponential potential.}
\begin{eqnarray}
A_S(k_*) &=& {2\over 5\sqrt{\pi}}
  \ \left( {H_*\over\mpl} \right)\ F\left({-n_T \over 2}\right)
  \ \sqrt{ {3-n\over 1-n}} \\
A_T(k_*) &=& {2\over 5\sqrt{\pi}}
  \ \left( {H_*\over\mpl} \right)\ F\left({-n_T\over 2}\right)
\end{eqnarray}
and the term coming from the small-argument expansion of the Hankel function
\begin{eqnarray}
F(x) & = & {1+2x\over 1+x}\,{2^x\,
        \Gamma({\textstyle{1\over2}}+x)\over\sqrt{\pi}} \\
     & = & 1 + (1-\gamma - \ln 2 )x + \cdots \\
     &\simeq&1 - 0.27x +\cdots
\end{eqnarray}
where $\gamma = 0.577\cdots$ is Euler's constant.  Further, $H_*$
is related to $V_*$ by
\begin{equation}
H_*^2 = {8\pi \over 3} {V_*\over 1-1/3p} = {8\pi V_*\over 3}\left[
        1 - n_T/6 + {\cal O}(n_T^2) \right]
\end{equation}

The consistency equation can be written as
\begin{equation}
T/S = -7\,n_T \ f_{T/S}^{({\rm PLI})}(\Ol,n_T)
\end{equation}
where for $k_*=H_0$ the correction factor is well fit by
\begin{equation}
f_{T/S}^{({\rm PLI})}(\Ol,n_T) = 0.97 + 0.58n_T + 0.25\Ol
    - (1+1.1n_T+0.28n_T^2 ) \Ol^2
\end{equation}
over the range of astrophysical interest: $0.8\le n<1$ and $0\le\Ol<0.8$.

Fitting the scalar + tensor CBR power spectrum to the
{\sl COBE} 2-year maps yields the normalization \cite{LidLytViaWhi}
\begin{equation}
A_S = \left( 2.25 \pm 0.2 \right) \times 10^{-5}
	   \ \Omega_0^{-0.775-0.04\ln\Omega_0}\,[\Omega_0/g(\Omega_0)]
           \ \exp\left[ 0.76\nt \right]
\end{equation}
valid over the same range of $n$ and $\Ol$ as above.
This allows us to calculate the one parameter to be determined, $H_*/\mpl$, as
a function of $\Ol$ and $n$; the results are shown in Fig.~3.

\section{Discussion}

Inflation makes three generic predictions: a flat Universe with nearly
scale-invariant spectra of scalar and tensor metric perturbations.
The anisotropy of the CBR offers a means of testing all three:  the positions
of the peaks or damping tail of the CBR anisotropy spectrum can test the
spatial flatness of the Universe \cite{flat} and measurements of the CBR power
spectrum can determine the relative amplitudes of scalar and tensor
perturbations and their spectral indices.
(In the case of tensor perturbations, unless $T/S>0.1$, only an upper limit
can obtained \cite{knoxmst}; and realistically, $n_T$ is likely to be
difficult to measure \cite{knox}.)
The CBR anisotropy probably offers the best means of measuring the scalar and
tensor metric perturbations and thereby constraining the properties of the
inflaton potential.

The scalar and tensor metric perturbations are both determined by
the underlying inflationary potential, and so conversely, knowledge of the
metric perturbations can be used to determine the potential and
its first few derivatives ($k-$space reconstruction).  To take
advantage of this in practice, one must relate first the metric perturbations
to CBR observables ($\ell$-space reconstruction).
However, doing this introduces
dependence upon cosmological parameters not associated with inflation
(the baryon density, the Hubble constant and a possible cosmological
constant).  As we have shown here, the most important of these
is the cosmological constant.

For almost a decade, the advantages of a cosmological constant
for inflationary cosmology have been touted -- accommodating
measurements of
the matter density which fall short of the critical density, lessening
the tension between measurements of the age and the Hubble constant,
large-scale structure which is in better agreement with that measured by
redshift surveys, and better agreement with
the baryonic content of clusters \cite{lcdm}.

In this paper we have carried out the program of perturbative
reconstruction, allowing for the possibility of
a cosmological constant.  In particular, at lowest order we
have derived the dependence upon a cosmological constant of the equations that
relate the observables $S$, $T$, $(n-1)$, and $n_T$ to the inflationary
potential and its first two derivatives.
We have done the same at second order, including the additional
observable $dn/d\ln k$ and the third derivative of the potential.
Likewise, we have also modified the consistency
relation to allow for a cosmological constant.
In addition, we have clarified previous notation/conventions and generalized
reconstruction to the use of other observables.  Now all that is needed
is a high-angular resolution map of the CBR sky!  With NASA considering
three proposals for a satellite mission in 1999 and ESA considering another
proposal, that could happen within the next five years or so.

\paragraph{Acknowledgments.}
We thank Andrew Liddle for several helpful conversations.
This work was supported in part by the DOE (at Chicago and
Fermilab) and the NASA (at Fermilab through grant NAG 5-2788).

\newpage


\begin{figure}
\begin{center}
\leavevmode
\epsfysize=12cm \epsfbox{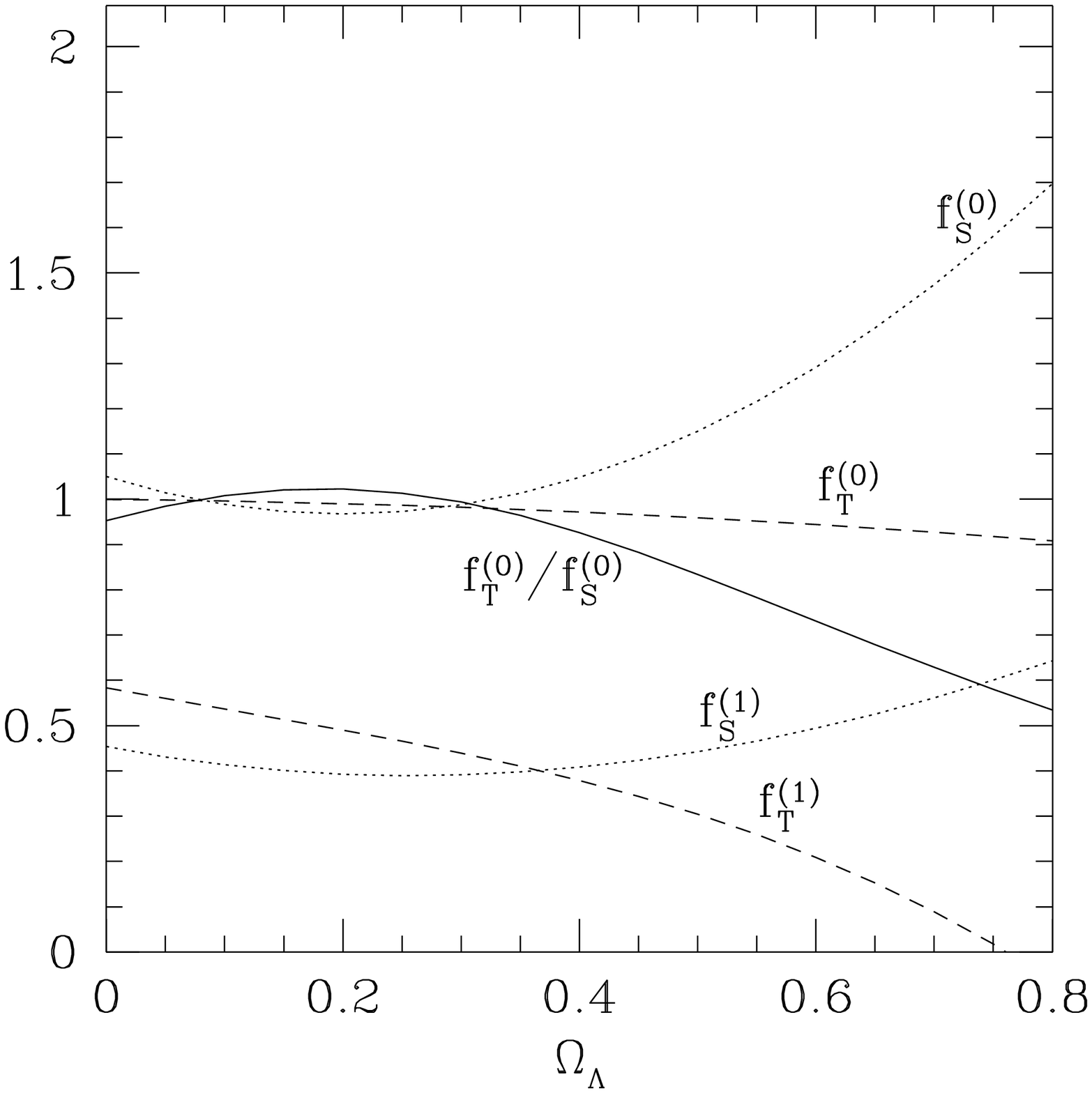}
\end{center}
\caption{The functions $f_i^{(0)}$ and $f_i^{(1)}$ and the ratio
$f_T^{(0)}/f_S^{(0)}$ as a function of $\Ol$.}
\end{figure}

\begin{figure}
\begin{center}
\leavevmode
\epsfysize=12cm \epsfbox{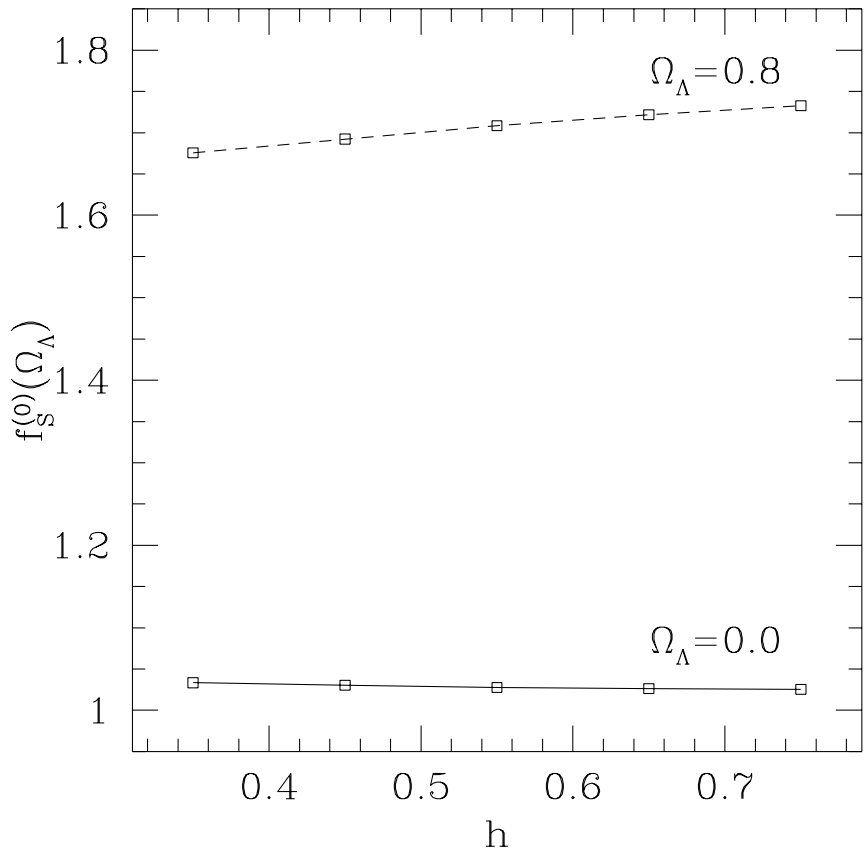}
\end{center}
\caption{The dependence of $f_S^{(0)}$ on the Hubble constant, for
$\Omega_\Lambda=0$ and $0.8$.  The dependence of $f_S^{(0)}$ on $h$ is
$\simlt 4\%$, much less than the $\Omega_\Lambda$ dependence.
Both dependences are due mostly to the evolution of the potentials from
last-scattering till the present (i.e.~the integrated Sachs-Wolfe effect).}
\end{figure}

\begin{figure}
\begin{center}
\leavevmode
\epsfysize=12cm \epsfbox{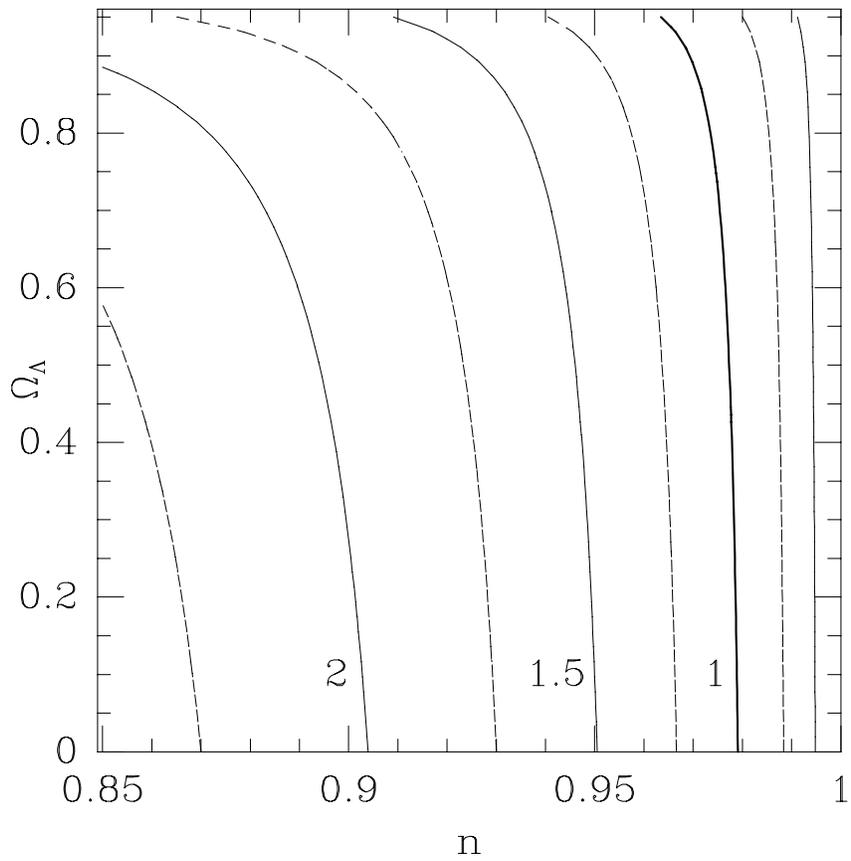}
\end{center}
\caption{The scale of power-law inflation, quantified by $10^5(H_*/\mpl)$,
as a function of $\Ol$ and $n$ ($H_*$ is the Hubble constant
during inflation when the scale $k_*=H_0$ crossed outside the horizon).
As described in Section 2.4, the {\sl COBE} normalization, $n$
and $\Omega_\Lambda$ fix $H_*/\mpl$.}
\end{figure}

\end{document}